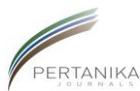

**MALAYSIAN JOURNAL OF MATHEMATICAL SCIENCES**

Journal homepage: http://einspem.upm.edu.my/journal

# On the Weak Localization Principle of the Eigenfunction Expansions of the Laplace-Beltrami Operator by Riesz Method

[1,2]**Anvarjon Ahmedov** and [2*]**Ahmad Fadly Nurullah Rasedee**

[1]*Department of Process and Food Engineering, Faculty of Engineering, Universiti Putra Malaysia, 43400 UPM Serdang, Selangor, Malaysia*

[2]*Institute for Mathematical Research, Universiti Putra Malaysia, 43400 UPM Serdang, Selangor, Malaysia*

E-mail: ahmadfadlynurullah@yahoo.com

*Corresponding author

## ABSTRACT

In this paper we deal with the problems of the weak localization of the eigenfunction expansions related to Laplace-Beltrami operator on unit sphere. The conditions for weak localization of Fourier-Laplace series are investigated by comparing the Riesz and Cesaro methods of summation for eigenfunction expansions of the Laplace-Beltrami operator. It is shown that the weak localization principle for the integrable functions $f(x)$ at the point $x$ depends not only on behavior of the function around $x$ but on the behavior of the function around diametrically opposite point $\overline{x}$.

## 1. INTRODUCTION

In this work, the problems of the weak localization of the eigenfunction expansions of Laplace-Beltrami operator on unit sphere are considered. It is stated in Riemann's localization theorem that the convergence or divergence for a one-dimensional Fourier series at a given point depends only on the behavior of the function $f \in L_1$ in an arbitrary small neighborhood of the point. The localization theorem may also formulated as follows: if two functions coincide in the neighborhood of a point, then the partial sums of their Fourier series have the same behavior there. The problems of localization of the eigenfunction expansions of the integrable functions are investigated in the papers by Ilin, 1968, Bonami and Clerc, 1973, Alimov, 1974, Pulatov, 1983, Topuria, 1987, Rakhimov, 2003 and Ashurov, 2012. For more reference on the problems of the convergence of the Fourier-Laplace series on unit sphere we refer the readers to Rakhimov, 2003. We note here that the investigation of the localization problems of the Fourier-Laplace series related to distributions on unit sphere was started by Rakhimov, 2003. He found the sufficient conditions for the localization of the Fourier-Laplace series of the distributions by Cesaro means in the classes of Nikolskii, Liouville and Sobolev.

In the work of Ahmedov, 2010, he proved the generalized principle of localization for the Fourier-Laplace series by Riesz Means of the order $s = (N-1)(1/p - 1/2)$, $1 \leq p \leq 2$. In the current work we investigate the weak localization of the eigenfunction expansions for the case of the Riesz means of the spectral expansions of the Laplace-Beltrami operator on unit sphere. We give positive answer to the following conjuncture: The sufficient conditions for localization can be weakened for the eigenfunction expansions of the Laplace-Beltrami operator on unit sphere?

## 2. EIGENFUNCTION EXPANSIONS OF THE LAPLACE-BELTRAMI OPERATOR

Let $S^N$ be a $N$ - dimensional unit sphere in $R^{N+1}$. We consider a symmetric and nonnegative elliptic operator $\Delta_s$, known as the Laplace-Beltrami operator. The Laplace-Beltrami operator $\Delta_s$ has in $L_2(S^N)$ a complete orthonormal system of eigenfunctions

$$\left\{ Y_1^{(k)}(x), Y_2^{(k)}(x), \ldots, Y_{a_k}^{(k)}(x) \right\}, \quad k = 0, 1, 2, \ldots$$

corresponding to the eigenvalues $\{\lambda_k = k(k+N-1)\}$ $k = 0, 1, 2, \ldots$ where $a_k$ is the multiplicity of the eigenvalue $\lambda_k$.

The Riesz means of the spectral expansions of the order $\alpha > 0$ related to the Laplace-Beltrami operator is defined by

$$E_n^\alpha f(x) = \sum_{k=0}^n \left(1 - \frac{\lambda_k}{\lambda_n}\right)^\alpha Y_k(f, x),$$

On the Weak Localization Principle of the Eigenfunction Expansions of the Laplace-Beltrami Operator by Riesz Method

Here

$$Y_k(f,x) = \sum_{j=1}^{a_k} Y_j^{(k)}(x) \int_{S^N} f(y) Y_j^{(k)}(y) d\sigma(y),$$

$$d\sigma(y) = \sin^{N-1}\phi_1 \sin^{N-2}\phi_2 \ldots \sin\phi_{N-1} d\phi_1 d\phi_2 \ldots d\phi_{N-1} d\varphi.$$

For any $x \in S^N$ the diametrically opposite of $x$ is denoted by $\bar{x}$. It is obvious that spherical distance $\gamma(x,\bar{x})$ between $x$ and $\bar{x}$ is equal to $\pi : \gamma(x,\bar{x}) = \pi$.

## 2. MAIN RESULTS

We proceed with the formulation of the main results of the paper.

**Theorem 1.** *Let* $f \in L_1(S^N)$. *If $f$ vanishes in a neighborhood of point $x \in S^N$ and $\bar{x} \in S^N$ then the Riesz means $E_n^\alpha f(x)$ of order $\alpha = \dfrac{N-1}{2}$ converges to zero at $x$:*

$$\lim E_n^{\frac{N-1}{2}} f(x) = 0.$$

The order $\alpha = \dfrac{N-1}{2}$ of the Riesz means $E_n^\alpha f(x)$ of the spectral expansions of the function $f$ is called critical order. It is proved that (see Stein, 1958) the condition $\alpha < \dfrac{N-1}{2}$ cannot guarantee convergence of the Riesz means $E_n^\alpha f(x)$ for integrable functions.

In the case of classes $L_p(S^N)$, $p > 1$ we have the following:

**Theorem .** *Let* $f \in L_p(S^N)$, $p > 2 - \dfrac{2}{N+1}$. *If $f$ vanishes in some neighborhood of $x \in S^N$ then the Riesz means $E_n^\alpha f(x)$ of order $\alpha \in \left[\dfrac{N-1}{2}, N-1\right)$ converges to zero at $x$:*

$$\lim_{n \to \infty} E_n^\alpha f(x) = 0.$$

# 3. PROOF OF MAIN RESULTS

## 3.1 Proof of Theorem 1

The Riesz means of the spectral expansions of the Laplace-Beltrami operator can be written as the following integral operator

$$E_n^\alpha f(x) = \int_{S^N} \Theta^\alpha(x, y, n) f(y)\, d\sigma(y)$$

with the kernel

$$\Theta^\alpha(x, y, n) = \sum_{k=0}^{n} \left(1 - \frac{\lambda_k}{\lambda_n}\right)^\alpha \sum_{j=1}^{a_k} Y_j^{(k)}(x) Y_j^{(k)}(y).$$

Using the Gegenbauer polynomials $P_k^{(\nu)}(t)$, by

$$P_k^{(\nu)}(t) = \frac{(-2)^k \Gamma(k+\nu)\Gamma(k+2\nu)}{\Gamma(\nu)\Gamma(2(k+\nu))} (1-t^2)^{-\left(\nu-\frac{1}{2}\right)} \frac{d^k}{dt^k}\left[(1-t^2)^{k+\left(\nu-\frac{1}{2}\right)}\right],$$

we can represent $\Theta^\alpha(x, y, n)$ in the following form

$$\Theta^\alpha(x, y, n) = \sum_{k=0}^{n} \left(1 - \frac{\lambda_k}{\lambda_n}\right)^\alpha \frac{\Gamma\left(k + \frac{N-1}{2}\right) \Gamma(k+N-1)}{\pi^{\frac{N+1}{2}}} P_k^{\left(\frac{N-1}{2}\right)}(\cos\gamma).$$

From this representation we can conclude that $\Theta^\alpha(x, y, n)$ depends only on the distance between $x$ and $y$. Due to this fact, we can denote the kernel as $\Theta^\alpha(x, y, n) = \Theta_n^\alpha(\cos\gamma)$.

**Theorem 3.** *Let $\Theta^\alpha(x, y, n)$ be the kernel of Riesz means of the spectral expansions.*

1. If $\left|\frac{\pi}{2} - \gamma(x, y)\right| < \frac{n}{n+1} \frac{\pi}{2}$, *as* $n \to \infty$, *then*

$$\Theta^\alpha(x, y, n) = n^{\frac{N-1}{2} - \alpha} (N-1) \frac{\sin\left[(n + N/2 + \alpha/2)\gamma - \pi(N-1+2\alpha)/4\right]}{(2\sin\gamma)^{\frac{N-1}{2}} (2\sin(\gamma/2))^{1+\alpha}}$$

$$+n^{\frac{N-3}{2}-\alpha}\frac{\eta_n(\gamma)}{(\sin\gamma)^{\frac{N+1}{2}}\left(\sin\frac{\gamma}{2}\right)^{1+\alpha}}+\frac{\varepsilon_n(\gamma)}{(n+1)\left(\sin\frac{\gamma}{2}\right)^{1+N}};$$

where $|\eta_n(\gamma)|<C$, $|\varepsilon_n(\gamma)|<C$;

2. If, $0\leq\gamma_0\leq\gamma\leq\pi$, $n>1$, then

$$\left|\Theta^\alpha(x,y,n)\right|<C_4 n^{N-1-\alpha},$$

3. If $0\leq\gamma\leq\pi$, $n>1$, then

$$\left|\Theta^\alpha(x,y,n)\right|=C_5 n^N.$$

For the proof, see Rasedee, 2015.

The integral over unit sphere of the function $g(x\cdot y)$, where $(x\cdot y)=$ scalar product, can be written as follows

$$\int_{S^{N-1}}g(x\cdot y)d\sigma(y)=\int_0^\pi\int_{(x,y)=\cos\gamma}g(\gamma)\,dt(y)d\gamma=\int_0^\pi g(\gamma)\,d\gamma\int_{(x,y)=\cos\gamma}f(y)\,dt(y)dt(y).$$

Using this representation we have

$$E_n^\alpha f(x)=\int_0^\pi \Theta_n^\alpha(\cos\gamma)\int_{(x,y)=\cos\gamma}f(y)\,dt(y)$$

here

$$\varphi(\gamma)=\int_{(x,y)=\cos\gamma}f(y)\,dt(y)$$

which yields

$$E_n^\alpha f(x)=\int_0^\pi \Theta_n^\alpha(\cos\gamma)\varphi(\gamma)d\gamma.$$

Let $f\in L_1(S^N)$ and $f(y)=0$, if $y\in B_{\varepsilon_1}(x)\cup B_{\varepsilon_2}(\overline{x})$, for some $\varepsilon_1,\varepsilon_2>0$. We then separate integration from $0$ to $\pi$ into the following integral

$$E_n^\alpha f(x)=\left(\int_0^{\varepsilon_1}+\int_{\varepsilon_1}^{\pi-\varepsilon_2}+\int_{\pi-\varepsilon_2}^\pi\right)\Theta_n^\alpha(\cos\gamma)\varphi(\gamma)d\gamma.$$

Using the fact that $f(y) = 0$ when $y \in B_{\varepsilon_1}(x) \cup B_{\varepsilon_2}(\bar{x})$

$$\int_0^{\varepsilon_1} \Theta_n^\alpha(\cos\gamma)\varphi(\gamma)d\gamma = \int_{\gamma(x,y)<\varepsilon_1} \Theta^\alpha(x,y,n)f(y)d\sigma(y)$$

$$= \int_{B_{\varepsilon_1}(x)} \Theta^\alpha(x,y,n)f(y)d\sigma(y) = 0$$

and

$$\int_{\pi-\varepsilon_2}^{\pi} \Theta_n^\alpha(\cos\gamma)\varphi(\gamma)d\gamma = \int_{\gamma(\bar{x},y)<\varepsilon_2} \Theta^\alpha(\bar{x},y,n)f(y)d\sigma(y)$$

$$= \int_{B_{\varepsilon_2}(\bar{x})} \Theta^\alpha(\bar{x},y,n)f(y)d\sigma(y) = 0$$

hence, allows us to rewrite the Riesz means of spectral expansions of $f$ as follows,

$$E_n^\alpha f(x) = \int_{\varepsilon_1}^{\pi-\varepsilon_2} \Theta_n^\alpha(\cos\gamma)\varphi(\gamma)d\gamma.$$

We estimate using Theorem 3, then have

$$E_n^\alpha f(x) = n^{\frac{N-1}{2}-\alpha}(N-1)\int_{\varepsilon_1}^{\pi-\varepsilon_2} \frac{\sin[(n+N/2+\alpha/2)\gamma - \pi(N-1+2\alpha)/4]\varphi(\gamma)}{(2\sin\gamma)^{\frac{N-1}{2}}(2\sin(\gamma/2))^{1+\alpha}}d\gamma$$

$$+ n^{\frac{N-3}{2}-\alpha}\int_{\varepsilon_1}^{\pi-\varepsilon_2} \frac{\eta_n(\gamma)\varphi(\gamma)}{(\sin\gamma)^{\frac{N+1}{2}}\left(\sin\frac{\gamma}{2}\right)^{1+\alpha}}d\gamma + \int_{\varepsilon_1}^{\pi-\varepsilon_2} \frac{\varepsilon_n(\gamma)\varphi(\gamma)}{(n+1)\left(\sin\frac{\gamma}{2}\right)^{1+N}}d\gamma = I_1 + I_2 + I_3.$$

Estimation of $I_3$.

Let $\alpha = \dfrac{N-1}{2}$, we see that $\dfrac{\varphi(\gamma)}{\sin\left(\dfrac{\gamma}{2}\right)^{N+1}}$ is integrable on $[\varepsilon_1, \pi - \varepsilon_2]$ for

On the Weak Localization Principle of the Eigenfunction Expansions of the Laplace-Beltrami Operator by Riesz Method

$\varepsilon_1, \varepsilon_2 > 0$, then it can be clearly seen that

$$|I_3| = \left| \int_{\varepsilon_1}^{\pi-\varepsilon_2} \frac{\varepsilon_n(\gamma)\varphi(\gamma)}{(n+1)\left(\sin\frac{\gamma}{2}\right)^{1+N}} d\gamma \right| \leq \frac{C}{n+1} \int_{\varepsilon_1}^{\pi-\varepsilon_2} \frac{|\varphi(\gamma)|}{\left(\sin\frac{\gamma}{2}\right)^{1+N}} d\gamma \to 0, \qquad n \to \infty.$$

Estimation of $I_2$.

Let $\alpha = \dfrac{N-1}{2}$. It is not difficult to see that

$$|I_2| = \frac{C}{n} \int_{\varepsilon_1}^{\pi-\varepsilon_2} \left| \frac{\varphi(\gamma)}{(\sin\gamma)^{\frac{N+1}{2}} \left(\sin\frac{\gamma}{2}\right)^{1+\alpha}} \right| d\gamma \to 0, \qquad n \to \infty.$$

when $\dfrac{\varphi(\gamma)}{(\sin\gamma)^{\frac{N+1}{2}}\left(\sin\frac{\gamma}{2}\right)^{1+\alpha}}$ is integrable on $[\varepsilon_1, \pi-\varepsilon_2]$ when $\varepsilon_1, \varepsilon_2 > 0$.

Estimation of $I_1$.

We have

$$I_1 = (N-1) \int_{\varepsilon_1}^{\pi-\varepsilon_2} \frac{\sin\left[\left(n+\frac{3N-1}{2}\right)\gamma - \frac{N-1}{2}\pi\right]\varphi(\gamma)}{(2\sin\gamma)^{\frac{N-1}{2}}(2\sin(\gamma/2))^{1+\alpha}} d\gamma,$$

which can be written as,

$$= \int_{\varepsilon_1}^{\pi-\varepsilon_2} \frac{(N-1)\cos\left(\frac{3N-1}{2}\gamma - \frac{N-1}{2}\pi\right)}{(2\sin\gamma)^{\frac{N-1}{2}}(2\sin(\gamma/2))^{1+\alpha}} \sin n\gamma \, d\gamma$$

$$+ \int_{\varepsilon_1}^{\pi-\varepsilon_2} \frac{(N-1)\sin\left(\frac{3N-1}{2}\gamma - \frac{N-1}{2}\pi\right)}{(2\sin\gamma)^{\frac{N-1}{2}}(2\sin(\gamma/2))^{1+\alpha}} \cos n\gamma \, d\gamma.$$

The functions

$$\frac{(N-1)\cos\left(\frac{3N-1}{2}\gamma - \frac{N-1}{2}\pi\right)}{(2\sin\gamma)^{\frac{N-1}{2}}(2\sin(\gamma/2))^{1+\alpha}}$$

and

$$\frac{(N-1)\sin\left(\frac{3N-1}{2}\gamma - \frac{N-1}{2}\pi\right)}{(2\sin\gamma)^{\frac{N-1}{2}}(2\sin(\gamma/2))^{1+\alpha}}$$

are integrable on $[\varepsilon_1, \pi - \varepsilon_2]$ for any $\varepsilon_1, \varepsilon_2 > 0$.

Then the Riemann-Lebesgue Theorem implies that

$$\int_{\varepsilon_1}^{\pi-\varepsilon_2} \frac{(N-1)\cos\left(\frac{3N-1}{2}\gamma - \frac{N-1}{2}\pi\right)}{(2\sin\gamma)^{\frac{N-1}{2}}(2\sin(\gamma/2))^{1+\alpha}} \sin n\gamma \, d\gamma \to 0,$$

$$\int_{\varepsilon_1}^{\pi-\varepsilon_2} \frac{(N-1)\sin\left(\frac{3N-1}{2}\gamma - \frac{N-1}{2}\pi\right)}{(2\sin\gamma)^{\frac{N-1}{2}}(2\sin(\gamma/2))^{1+\alpha}} \cos n\gamma \, d\gamma \to 0,$$

as Fourier coefficients.

### 3.2 Proof of Theorem 2

From

$$\left|E_n^\alpha f(x)\right| \le Cn^{\frac{N-1}{2}-\alpha} \int_{\pi-\varepsilon}^{\pi} \frac{|\varphi(\gamma)|}{(\sin\gamma)^{\frac{N-1}{2}}} \, d\gamma,$$

we obtain then $E_n^\alpha f(x) \to 0$ for $\alpha > \frac{N-1}{2}$. If $\alpha = \frac{N-1}{2}$ then using the Lebesgue Theorem it can be shown that the integral

$$\int_{\pi-\varepsilon}^{\pi} \frac{|\varphi(\gamma)|}{(\sin\gamma)^{\frac{N-1}{2}}} \, d\gamma$$

can be made less than any $\varepsilon > 0$.

If $\alpha > \dfrac{N-1}{2}$ we have

$$|E_n^\alpha f(x)| \le C \int_{\pi-\varepsilon}^{\pi} \frac{|\varphi(\gamma)|}{(\sin\gamma)^{\frac{N-1}{2}}} \left( \frac{n^{\frac{N-1}{2}-\alpha}}{\left(\sin\frac{\gamma}{2}\right)^{1+\alpha}} + \frac{1}{n\left(\sin\frac{\gamma}{2}\right)^{\frac{N-1}{2}+2}} \right) d\gamma$$

$$\le C n^{\frac{N-1}{2}-\alpha} \int_{\pi-\varepsilon}^{\pi} \frac{|\varphi(\gamma)|}{(\sin\gamma)^{\frac{N-1}{2}}} d\gamma.$$

We will show that the integral

$$\int_{\pi-\varepsilon}^{\pi} \frac{|\varphi(\gamma)|}{(\sin\gamma)^{\frac{N-1}{2}}} d\gamma$$

is finite.

Let $p > 1$ satisfy $\dfrac{1}{p}+\dfrac{1}{q}=1$. Using Holder inequality we obtain

$$\int_{\pi-\varepsilon}^{\pi} \frac{|\varphi(\gamma)|}{(\sin\gamma)^{\frac{N-1}{2}}} d\gamma = \int_{\pi-\varepsilon}^{\pi} \frac{|\varphi(\gamma)|}{(\sin\gamma)^{\frac{N-1}{q}}} \frac{1}{(\sin\gamma)^{\frac{N-1}{2}-\frac{N-1}{q}}} d\gamma$$

$$\le C \left\{ \int_{\pi-\varepsilon}^{\pi} \frac{|\varphi(\gamma)|^p}{\sin^{\frac{(N-1)p}{q}}\gamma} d\gamma \right\}^{\frac{1}{p}} \left\{ \int_{\pi-\varepsilon}^{\pi} \frac{d\gamma}{(\sin\gamma)^{\frac{(N-1)q}{2}-N-1}} \right\}^{\frac{1}{q}}$$

$$\le C \left\{ \int_{\pi-\varepsilon}^{\pi} \frac{\int_{(x,y)=\cos\gamma} |f(y)|^p |\gamma^{N-2}|^{\frac{p}{q}} dt(y)}{(\sin\gamma)^{\frac{(N-1)p}{q}}} \right\}^{\frac{1}{p}} \left\{ \int_0^{\varepsilon} \frac{d\gamma}{(\sin\gamma)^{\frac{(N-1)q}{2}-N-1}} \right\}^{\frac{1}{q}}$$

$$\le C \left\{ \int_{S^N} |f(y)|^p \, dS(y) \right\}^{\frac{1}{p}} \left\{ \int_0^{\varepsilon} \frac{d\gamma}{(\sin\gamma)^{\frac{(N-1)q}{2}-N-1}} \right\}^{\frac{1}{q}}$$

The first integral is finite because $f \in L_p(S^N)$, and if we choose $p > 2 - \dfrac{2}{N+1}$, then $q$ satisfies the following condition: $q\dfrac{N-1}{2} - N < 2$. By similar method we can show that the second integral is also finite.

## CONCLUSION

In the current work, sufficient conditions for weak localization are established in the classes of integrable functions. It is shown that the convergent of the Fourier Laplace series of the integrable functions at one point requires investigations of the behavior of the function not only in the given point, but also its diametrical opposite.

## ACKNOWLEDGEMENT

This paper has been supported by Ministry of Higher Education (Mohe) for its MyPhd sponsorship and Universiti Putra Malaysia under Research University Grant (RUGS), Project number is 05-01-12-1630RU.